# 'Partisan Bias' is Like 'Cancer'

Alec Ramsay, Dave's Redistricting[*]

## Abstract

The colloquial phrase "partisan bias" encompasses multiple distinct conceptions of bias, including partisan advantage, packing & cracking, and partisan symmetry. All are useful and have their place, and there are several proposed measures of each. While different measures frequently signal the direction of bias consistently for redistricting plans, sometimes the signals are contradictory: for example, one metric says a map is biased towards Democrats while another metric say the same map is biased towards Republicans.[1] This happens most frequently with metrics that measure different kinds of bias, but it also occurs between measures in the same category. These inconsistencies are most pronounced in states where one party is dominant, but they also occur across the full range of partisan balance. The political geography of states also influences the frequency with which various measures are inconsistent in their assessment of bias. No subset of metrics is always internally consistent in their signal of bias.

Word count: 4,156 words, excluding references.

## 1 – Introduction

To say that someone has "cancer" is to say that they have a general kind of disease (i.e., cancer) without saying much specific about their individual condition (e.g., prostate cancer). There are hundreds of different types of cancer. The technical linguistic concept for terms like "cancer" that have different context-dependent meanings at different levels of abstraction is that they are multiordinal.[2]

In redistricting there are lots of metrics that purport to measure partisan bias, but discussion of the degree to which a set of districts favors one party or another, i.e., is biased, is muddied because "partisan bias" is similarly multiordinal. It is an abstract umbrella term that encompasses at least three distinct types of bias.

---

[*] I work on Dave's Redistricting (DRA) in Seattle, Washington, USA and do independent research. I would like to thank Jon Eguia, Gary King, Nick Stephanopoulos, and Greg Warrington for their feedback on early drafts. I would also like to specially acknowledge DRA, which I use to show example maps here, and the Voting and Election Science Team (VEST) and the Redistricting Data Hub (RDH) which supplied most of the election data in DRA.

[1] I am intentionally fuzzing the difference between "parties" and "voters" here.
[2] As a cancer survivor myself, I feel comfortable using this analogy.

To help sharpen our understanding of the similarities and differences between the several types and many measures of partisan bias, I classify eight prominent metrics into three categories—partisan advantage, packing & cracking, and partisan symmetry—and compare their signals of bias for millions of plans across multiple states and legislative chambers from a recent study [TPR25].

## 2 – Previous Work

Proponents of individual measures of partisan bias have discussed the type of bias that their proposed metric purports to measure, why it is important to understand that kind of bias, and have contrasted them with some other measures [for example SM15, MB15, Wan16, War17, and KKR20].

Warrington compared several metrics—including his declination ($\delta$) measure, the efficiency gap ($EG$), geometric seat bias ($\beta$), mean-median difference ($mM$), and lopsided outcomes ($LO$)—through the lens of packing & cracking, the gerrymandering technique that results in votes in different districts having unequal weights [War19]. He defined good measures of packing & cracking as satisfying two postulates. He analyzed how the metrics he studied performed over a large set of historical elections well as a dozen carefully crafted hypothetical elections and concluded that $\delta$ was the best measure from this perspective.

Katz *et al* analyzed many of the same metrics—(dis)proportionality ($PR$), mean-median difference ($mM$), lopsided outcomes ($LO$), declination ($\delta$), the efficiency gap ($EG$), and geometric seat bias ($\beta$)—but mathematically and through a different lens: seats-denominated partisan symmetry which they formally defined [KKR20]. They showed that of those metrics only $\beta$ is a valid estimator of that quantity of interest.

John Nagle and I studied all these metrics and a few others—in particular, seat bias ($\propto_S$) and vote bias ($\propto_V$) which are widely discussed, e.g. [MB15]—with a particular focus on assessing partisan bias in states where one party is dominant [NR21]. We studied eleven 2012 congressional plans.

I defined a new classification of metrics that measure partisan bias as the difference between seats won[3] and some normative expectation of the relationship between votes and seats [Ram23]. I enumerated five properties that valid measures of this seats-votes partisan advantage concept ($PA|SV$) must satisfy [SM15].[4] The previously defined efficiency gap ($EG$) is a measure of partisan advantage. This third category of partisan bias is distinct from the previous two.

---

[3] Or likely to be won.

[4] There are other benchmarks for partisan advantage such as Jon Eguia's jurisdictional advantage [Egu22] among others [BE24].

The ALARM project produced ensembles for all states with two or more congressional districts [MKS+22]. They used Sequential Monte Carlo (SMC) to produce ensembles of 5,000 plans and calculated the efficiency gap ($EG$) and geometric seats bias ($\beta$) for those plans. Cannon *et al* analyzed problematic redundancy in redistricting ensembles produced using SMC [CDD24].

Two colleagues and I recently studied parameter effects in ensembles generated using the Markov chain Monte Carlo Recombination algorithm (ReCom) [TPR25]. We generated statistically sound ensembles with 20,000 subsampled plans for congressional and state upper & lower house redistricting plans for seven states (NC, MI, NY, IL, FL, OH, and WI), and we computed a wide array of metrics for each plan. We explored how various parameter settings influenced the resulting scores. That dataset is the basis of this paper.

## 3 – Contributions

To the best of my knowledge, no one has yet compared common measures of partisan bias over large statistically sound samples of redistricting plans that systematically explore the political geography of multiple states or studied the three subcategories of partisan bias—partisan advantage, packing & cracking, and partisan symmetry—separately and then in contrast. That is the main contribution of this paper.

As part of doing that, I make a few other contributions:

1. I diagnose the root problem in discourse about partisan bias in redistricting: multiordinality.
2. I elevate and reinforce subcategories of partisan bias metrics—partisan advantage, packing & cracking, and partisan symmetry—that I previously noted in the context of defining partisan advantage [Ram23]. I define them descriptively and analytically.
3. I provide concrete examples of different measures conflicting in their signal of bias and explain how and why this occurs.
4. I show that no pair of metrics is always consistent in their signal of bias.

I sketch my methodology in the next section and follow it with two sections of results.

## 4 – Terminology, Scope, and Methodology

First and foremost, I use "partisan bias" throughout to refer to the general category of metrics that measure some aspect of how fair/unfair a redistricting map is from some partisan perspective and use "geometric seat bias" to refer to the specific measure defined in [KKR20] which they also called "partisan bias."

I study eight metrics that fall into three subcategories of partisan bias: (dis)proportionality ($PR$) and the efficiency gap ($EG$) are measures of partisan advantage; declination ($\delta$), mean-median difference ($mM$), and lopsided outcomes ($LO$) are measures of packing & cracking; and geometric seat bias ($\beta$), seat bias ($\propto_S$), and vote bias ($\propto_V$) are measures of partisan symmetry. All are functionally defined in detail in §4 of [Ram23].

I use the scores from [TPR25] noted above. There we produced 315 ensembles: seven states—NC, MI, NY, IL, FL, OH, and WI; three chambers—Congress and state upper and lower houses; and 15 ReCom variants (defined in [TPR25, §4.4]). I exclude the $A_1 \ldots A_4$ statistical variant ensembles from my analysis here, leaving 11 ensembles. Each has 20,000 plans subsampled from ReCom chains 50 million long, except reversible ReCom which used a chain of one billion. Altogether, there are 4,620,000 plans in this subset (hereafter "the sample").

To highlight the differences between the metrics in different categories of partisan bias, as well as show the similarities between metrics within each category, I look at whether the metrics indicate that a plan favors Republicans or Democrats, + signed values or − signed values, respectively by convention. Two metrics agree on the whether a plan favors one party or the other if their signs are the same for that plan; the metrics contradict each other, if the signs are different.[5]

I first explore conflicting signals of bias *within* each of the three categories (§5). Then I look at conflicting signals *between* internally consistent categories of metrics (§6).

## 5 – Conflicts Within Categories

In this section I examine conflicting signals of partisan bias within each of the three categories of metrics: partisan advantage (§5.1), packing & cracking (§5.2), and partisan symmetry (§5.3), respectively.

### 5.1 – Conflicts Within Partisan Advantage

While there are many possible ideals for how votes should be translated into seats in a measure of seats-votes partisan advantage ($PA|SV$), there are two main measures: (dis)proportionality ($PR$) and the efficiency gap ($EG$) [Ram23]. They embody different normative expectations about seat shares given vote shares: $PR$ measures the degree to which vote shares are translated into seat shares proportionally, a responsiveness of one, while $EG$ incorporates a responsiveness of two. Both are functions of just two variables: statewide vote share ($\overline{V}$) and statewide seat share ($\overline{S}$) and measure "whether a districting arrangement causes **harm to a party** because it wins fewer seats than its 'fair' share" [MB15, p. 312, emphasis added].

These two metrics give conflicting signals about which party a redistricting plan favors, when the majority party wins more than a proportional share of seats but that share is less than the two times winner's bonus baked into $EG$.[6]

---

[5] One metric, declination ($\delta$) can sometimes be undefined which is treated as matching the sign for the other metric.
[6] These cases fall in the D2 region of [Ram23, Figure 1].

Table 1 shows that $PR$ and $EG$ yield conflicting signals of bias across all seven states in the sample (165 of 231 ensembles).[7] The table is sorted by absolute partisan lean above/below 50%, and the righthand column shows the percentage of plans for which the metrics gave conflicting signals of bias.

| State | Partisan Lean | Conflicts |
|-------|---------------|-----------|
| NC | 0.57% | 1.04% |
| WI | 0.68% | 0.35% |
| FL | 1.63% | 2.94% |
| MI | 1.88% | 2.06% |
| OH | 3.62% | 0.74% |
| IL | 8.17% | 71.68% |
| NY | 14.78% | 24.33% |

Table 1: Partisan Advantage

As the table shows, the conflicting signals are heavily concentrated in Illinois and New York, both heavily unbalanced politically. While the conflicts are more likely and more frequent when a state is unbalanced politically, the table also shows that conflicts occur across the full range of partisan lean and the conflict rate isn't a simple function of it. In other words, the political geography of a state influences the characteristics of the plans in ensembles generated with neutral criteria[8] and, thus, how these two metrics will differentially assess bias.

Figure 1 shows a sample Illinois congressional plan with this divergence (map: bit.ly/4gy74zP).[9] The inferred seats-votes curve runs right between the two normative lines: $PR$ is -4.86% because the expected seat share is more than proportional (dotted line) but $EG$ is 3.31% because it is below a two-times winner's bonus (dashed line). Interpreting $EG$ in this region can be tricky: here Democrats are likely to win more than a proportional share of seats, suggesting the plan favors them by $PR$, but less than the two-times winner's bonus of $EG$, so it somewhat counterintuitively signals that the plan favors Republicans.

---

[7] 231 = 7 x 3 x 11 state, chamber, and ensemble combinations
[8] Meaning that only neutral criteria, such as compactness and county splitting, were used by the ensemble generating algorithm.
[9] Plan IL / Congress / A0 / 000002500 in the sample.

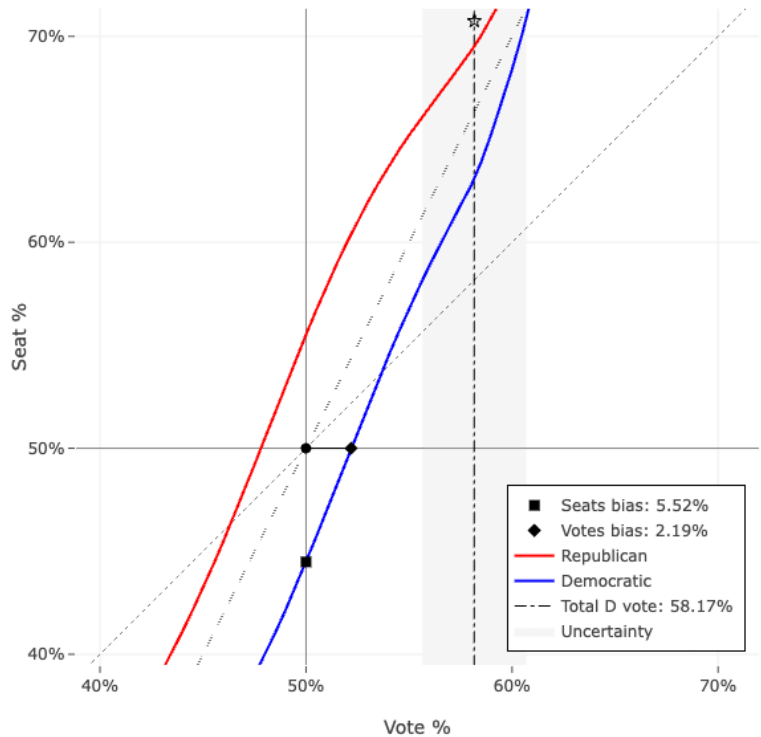

Figure 1: $EG$ vs. $PR$

## 5.2 – Conflicts Within Packing & Cracking

Packing & cracking partisan voters is the main technique used to create partisan gerrymanders [MB15, p. 3, War17, p.1]. In contrast to measures of partisan advantage, the three measures of packing & cracking here—declination ($\delta$), mean-median ($mM$),[10] and lopsided outcomes ($LO$)—depend on vote shares by district: ($v = v_1 \ldots v_N$) and ask "whether a districting arrangement causes an **unequal weighting of votes** for one set of partisan **voters** versus the other" [MB15, p. 312, emphasis added]. In other words, these measures are designed to gauge the degree to which the votes for voters in different districts are weighted (un)equally, i.e., the degree to which a plan has a voting rights issue [McD09, p. 2].

Table 2 shows the frequency with which these three measures give conflicting signals about bias, i.e., when some indicate a plan is biased in favor of one party, but others indicate the opposite. The measures conflict in all seven states in the sample (181 of 231 ensembles).[11] To further elucidate differences between these metrics, the pairwise conflicts are shown in the three righthand columns.

---

[10] I use the average district vote share as opposed to the statewide vote share formulation.
[11] Declination is undefined in 75,736 of 200,000 plans (37.87%) in 10 of 11 ensembles for NY/Congress. As undefined, they are not considered inconsistent.

| State | Partisan Lean | Conflicts | $\delta\|MM$ | $\delta\|LO$ | $LO\|MM$ |
|-------|---------------|-----------|--------------|--------------|----------|
| NC | 0.57% | 11.79% | 7.16% | 8.51% | 7.82% |
| WI | 0.68% | 0.47% | 0.35% | 0.13% | 0.47% |
| FL | 1.63% | 18.74% | 5.63% | 17.94% | 13.82% |
| MI | 1.88% | 0.32% | 0.32% | 0.01% | 0.32% |
| OH | 3.62% | 70.08% | 15.59% | 68.01% | 56.46% |
| IL | 8.17% | 4.98% | 4.98% | 4.90% | 0.08% |
| NY | 14.78% | 0.02% | 0.02% | 0.02% | 0.00% |

Table 2: Packing & Cracking

Even though the three metrics all purport to gauge the degree of packing & cracking in a redistricting plan, they sometimes disagree as to which party's voters are harmed. Sometimes in some states the rate of these differences is significant—e.g., North Carolina and Florida. Sometimes the conflicting signals are almost universal—e.g., Ohio.

This map (bit.ly/4gCbQwf) is an example for Ohio where $mM$ and $\delta$ indicate a small Republican bias (0.75% and 0.18°, respectively), while $LO$ indicates a Democratic bias (-3.72%).[12] As previously noted, the characteristics of a state's political geography influence the plans in a neutral ensemble and the resulting measurements of partisan bias.

## 5.3 – Conflicts Within Partisan Symmetry

Partisan symmetry is the principle that a redistricting plan should treat the two parties equally. In contrast to measures of partisan advantage, partisan symmetry says it's fine for one party to win more seats than expected given the share of votes received, provided the other party would also win that number of seats for the same vote share in that counterfactual.

There are three main measures of partisan symmetry in common use. Katz *et al* formally defined seats-denominated partisan bias ($\beta$) [KKR20], which, as noted above, I call geometric seats bias throughout for clarity. While not satisfying their formal definition of partisan bias, seat bias ($\propto_S$) and vote bias ($\propto_V$) are also widely considered measures of partisan symmetry (see [MB15] and [NR21], for example). While both $\beta$ and $\propto_S$ are seats denominated, I include votes-denominated $\propto_V$ here because it is a simple function of $\propto_S$ and the swing ratio [MB15, p. 314]. All three metrics are derived from an inferred seats-votes curve which is a function of both the statewide vote share ($\bar{V}$) and vote shares by district: ($v = v_1 \dots v_N$).

Table 3 shows that these measures sometimes conflict in all seven states in the sample (132 of 231 ensembles).

---

[12] Plan OH / Congress / R25 / 047750000 in the sample.

| State | Partisan Lean | Conflicts |
|-------|--------------:|----------:|
| NC | 0.57% | 0.03% |
| WI | 0.68% | 0.01% |
| FL | 1.63% | 3.60% |
| MI | 1.88% | 0.01% |
| OH | 3.62% | 2.73% |
| IL | 8.17% | 0.02% |
| NY | 14.78% | 0.14% |

Table 3: Partisan Symmetry

Conceptually, $\propto_S$ and $\propto_V$ can never conflict. When an inferred seats-votes curve is asymmetric—i.e., it does not pass through the center point of symmetry at $(0.5, 0.5)$—it either passes in the lower right quadrant or the upper left quadrant meaning that the signs for the two are necessarily the same.[13]

That leaves the case when these two both indicate bias in favor of one party but β signals bias in favor of the other party. As the table shows, this occurs infrequently in these ensembles, but it does happen. For example, in this New York congressional plan (map: bit.ly/47S2lqv) $\propto_S$ and $\propto_V$ indicate Democratic bias (-1.31% and -0.53%, respectively) but β indicates the opposite (3.69%).[14]

# 6 – Conflicts Between Categories

In this section I examine the frequency with each pair of *categories* give internally consistent signals of bias but nonetheless give conflicting signals of bias from each other.

## 6.1 – Partisan Advantage / Packing & Cracking Conflicts

Table 4 shows the frequency with which internally consistent measures of partisan advantage and packing & cracking nevertheless conflict in their respective signals of bias. Conflicts occur in six of the seven states in the sample (135 of 231 ensembles).

---

[13] Occasionally in the sample $\propto_S$ has a very small negative value while $\propto_V$ has a value of negative zero due to Python floating point behavior.
[14] Plan NY / Congress / A0 / 014170000 in the sample.

| State | Partisan Lean | Conflicts |
|-------|--------------:|----------:|
| NC | 0.57% | 1.50% |
| WI | 0.68% | 0.19% |
| FL | 1.63% | 0.00% |
| MI | 1.88% | 0.48% |
| OH | 3.62% | 5.23% |
| IL | 8.17% | 23.41% |
| NY | 14.78% | 75.65% |

Table 4: Partisan Advantage / Packing & Cracking

The reliability of $mM$ and $\delta$ have been previously explored with respect to $EG$, especially in states that are unbalanced politically, where "reliability" meant "consistent with $EG$."[15] The results here, reinforce that view and extend it to the $LO$ measure and a $PR$ baseline.

The table also shows that the conflicting signals aren't a simple function of the partisan lean of a state. Again, in addition to partisan lean, the political geography of a state influences the characteristics of the plans in ensembles generated with neutral criteria and, thus, the resulting measurements of partisan bias.

An example illustrates how this happens, most commonly in states with a dominant political party. Typical voting in New York state splits heavily Democratic, roughly 65% Democratic and 35% Republican. This congressional map (bit.ly/4pynr3i) is quite disproportional in favor of D's (18.70%)—they likely win 5 of 26 seats *more* than expected given their vote share.[16] But all three packing & cracking measures report the map as having a Republican-favorable bias: $mM$ is 7.37%, $LO$ is 20.38%, and $\delta$ is 25.95°. This is because Democratic-leaning voters are unavoidably packed in many districts. There are simply too many Democrats to *not* pack them. The packing & cracking metrics indicate that the weighting of the votes for the Democratic *voters* is less than that for Republican voters, even though the Democratic *party* wins more than an expected number of seats.[17]

Measures of packing & cracking and measures of partisan advantage probe different aspects of redistricting plans: voting rights and representational rights, respectively. As these concepts of partisan bias are functions of different abstractions of plans, it's not surprising that they would sometimes if not frequently yield different signals of bias.

---

[15] See [Pla] at https://planscore.org/#!2022-ushouse: "All of these metrics are reliable when a state is competitive, but only the efficiency gap and declination should be trusted when one party predominates in a state." with respect to Partisan Bias, Mean-Median, and Declination and [SM18, p. 1560].

[16] Plan NY / Congress / Pop- / 048322500 in the sample.

[17] As this example shows (map: bit.ly/3VXN4gn), $mM$ and $LO$ can even report a plan as having a bias in favor of the minority party when the majority party is likely to win *all* districts! When one party "sweeps" all the seats, $\delta$ is undefined.

## 6.2 – Partisan Advantage / Partisan Symmetry Conflicts

Measures of partisan advantage and measures of partisan symmetry can also each be internally consistent for a plan and yet give conflicting signals of bias.

To understand this how this can happen, you need to understand seats-votes curves. They are essentially S-shaped, and when they are symmetric, they pass through a center point of symmetry at (0.5,0.5). As a result, the more *a*symmetric an inferred seats-votes curve is and the more unbalanced a state is politically, the more likely it is that the curve passes by the center point of symmetry on one side of the 45° line of proportionality (where $S = V$) but passes through the actual $(S, V)$ point—i.e, the point where one party gets a large majority of the votes and an even larger share of the seats—on other side of it. In these cases, $\propto_S$ and $\propto_V$ indicate that the plan is biased in favor of one party even though the other party wins more than a proportional share of seats, possibly even more than implied by the two-times winner's bonus of $EG$.

This phenomenon is illustrated in Figures 1 and 2 (reproduced from [Ram23]), showing a Democratic-favorable plan in Illinois in terms of seats won having Republican-favorable $\propto_S$ and $\propto_V$ and the opposite for the Texas plan, respectively.

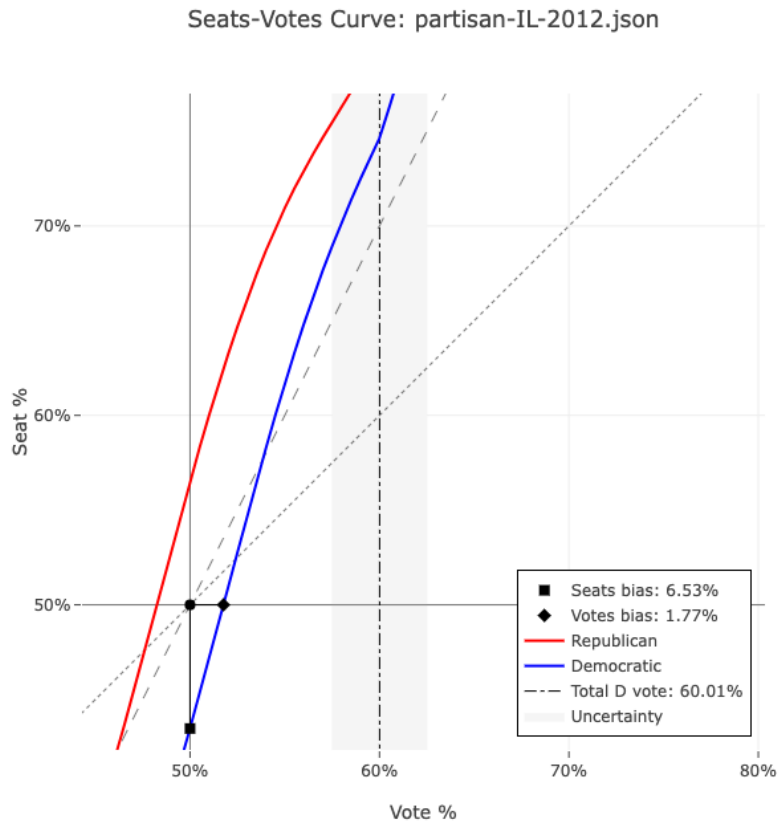

Figure 1: IL 2012 Congressional

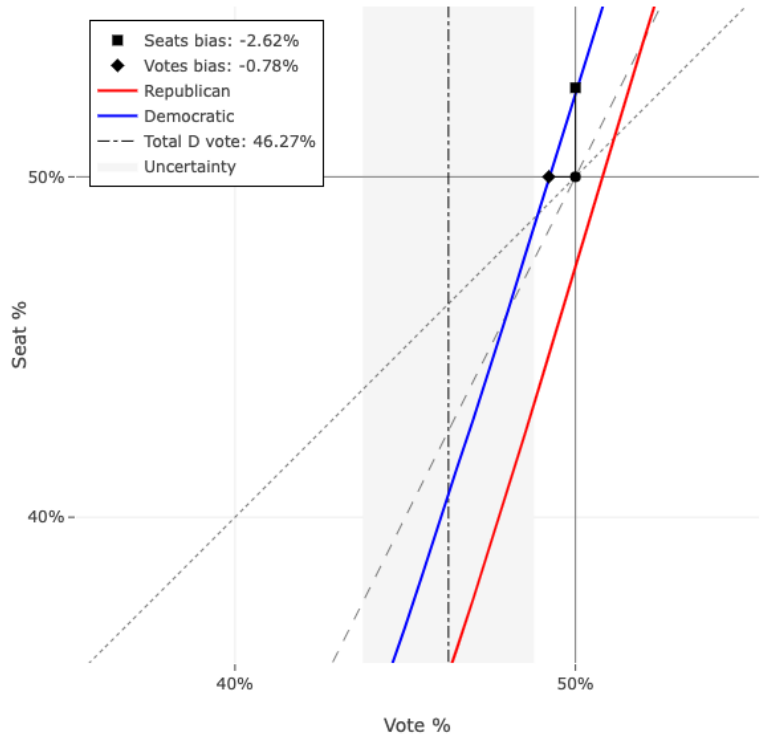

Figure 2: TX 2020 Congressional

Table 5 shows that these conflicts occur in all seven states in the sample (170 of 231 ensembles).

| State | Partisan Lean | Conflicts |
|-------|--------------|-----------|
| NC | 0.57% | 1.83% |
| WI | 0.68% | 0.22% |
| FL | 1.63% | 11.55% |
| MI | 1.88% | 0.61% |
| OH | 3.62% | 7.11% |
| IL | 8.17% | 28.31% |
| NY | 14.78% | 75.61% |

Table 5: Partisan Advantage / Partisan Symmetry

The unreliability of $\beta$ in unbalanced states has previously been explored, where "reliability" meant "consistent with $EG$" (see Footnote 15). As the table shows though, $\beta$ for New York must almost *always* contradict the signal of bias from measures of partisan advantage. The rate of conflict for Illinois is also quite high.

The table also illustrates or implies three other less well understood phenomena. First, $\propto_S$ and $\propto_V$ must also be as unreliable as $\beta$ in politically unbalanced states. Second, there isn't a smooth relationship between partisan lean and the rate of conflicting signals. For example, while Florida is fairly balanced politically, internally consistent measures of partisan symmetry conflict with internally consistent measures of partisan advantage a significant percentage of the time. Finally, even in almost balanced states, like North Carolina, internally consistent measures of partisan symmetry can still produce counterintuitive results from the standpoint of either measure partisan advantage.

Beyond just giving contradictory signals though, the *magnitude* of these differences can also be quite large. For example, $\beta$, $\propto_S$, and $\propto_V$ report this map (bit.ly/4pDojEa) as having a small Republican-favorable bias (1.06%, 1.86%, and 1.28%, respectively) even though the plan is quite disproportional in favor of Democrats (24.66%).[18]

As before, measures of partisan symmetry and measures of partisan advantage probe different aspects of redistricting plans: here whether plans treat parties equally as opposed to whether a party wins more (or less) seats than expected given their vote share. As the underlying concepts of partisan bias are functions of different abstractions of plans, it's again not surprising that they would sometimes if not frequently produce different signals of bias.

## 6.3 – Packing & Cracking / Partisan Symmetry Conflicts

Table 6 shows the results of comparing the signals of bias from packing & cracking metrics, on the one hand, and measures of partisan symmetry, on the other. Conflicts occur in five of the seven states in the sample (96 of 231 ensembles).[19]

| State | Partisan Lean | Conflicts |
|-------|--------------|-----------|
| NC | 0.57% | 0.21% |
| WI | 0.68% | 0.08% |
| FL | 1.63% | 2.02% |
| MI | 1.88% | 0.00% |
| OH | 3.62% | 0.61% |
| IL | 8.17% | 0.00% |
| NY | 14.78% | 0.00% |

Table 6: Packing & Cracking / Partisan Symmetry

While the two categories can sometimes conflict in their signals of bias, it is relatively rare: in the sample, when packing & cracking and partisan symmetry metrics are each internally consistent, they almost always agree on the signal of bias. This is an example (map: bit.ly/3KdMubQ) when that is <u>not</u> true.[20]

---

[18] Plan NY / Congress / D / 031252500 in the sample.

[19] There are a few conflicts in Michigan, but the rate is extremely low and rounded away.

[20] Plan FL / Congress / A0 / 010995000 in the sample.

## 7 – Pairwise Conflicts of Metrics

The analyses above explores the state-by-state rate of conflicting signals coming from metrics within each category (§5) and between categories of metrics (§6). It illustrates how partisan balance and political geography influence those rates of conflict. Table 7 below shows the overall rate of conflict between pairs of metrics in the sample regardless of state, chamber, and ensemble variant.

|  | $PR$ | $EG$ | $mM$ | $LO$ | $\delta$ | $\propto_S$ | $\propto_V$ | $\beta$ |
|---|---|---|---|---|---|---|---|---|
| $PR$ | - | 14.73% | 33.90% | 44.20% | 28.35% | 32.70% | 32.67% | 33.46% |
| $EG$ | 14.73% | - | 19.56% | 29.39% | 14.25% | 17.93% | 17.90% | 18.65% |
| $mM$ | 33.90% | 19.56% | - | 11.28% | 4.86% | 4.59% | 4.56% | 5.21% |
| $LO$ | 44.20% | 29.39% | 11.28% | - | 14.22% | 12.09% | 12.06% | 11.87% |
| $\delta$ | 28.35% | 14.25% | 4.86% | 14.22% | - | 3.54% | 3.50% | 4.00% |
| $\propto_S$ | 32.70% | 17.93% | 4.59% | 12.09% | 3.54% | - | 0.00% | 0.93% |
| $\propto_V$ | 32.67% | 17.90% | 4.56% | 12.06% | 3.50% | 0.00% | - | 0.90% |
| $\beta$ | 33.46% | 18.65% | 5.21% | 11.87% | 4.00% | 0.93% | 0.90% | - |

Table 7: Pairwise Conflicts

The three boxes along the diagonal aggregate the results of Tables 1-3, while the two sets of mirror-image boxes below and above the diagonal aggregate the results of Tables 4-6.

The table shows that the sample includes plans for which every pair of metrics conflict in their signals of bias, except $\propto_S$ and $\propto_V$ which can't conflict (as noted in §5.3).

A subset of these results is noteworthy. It has previously been asserted that $\propto_S$, $\propto_V$, and $\delta$ and two other measures not considered here—global symmetry and gamma—"provide mutually consistent values in all states, thereby providing a core of usable measures [of partisan bias] for unbalanced states" which can be combined into a composite metric dubbed $\Omega$ [NR21, P. 116]. That study looked at the official redistricting plan from each of 11 states.

| State | Partisan Lean | Diff Rate |
|---|---|---|
| NC | 0.57% | 0.65% |
| WI | 0.68% | 0.24% |
| FL | 1.63% | 14.02% |
| MI | 1.88% | 0.02% |
| OH | 3.62% | 4.75% |
| IL | 8.17% | 4.91% |
| NY | 14.78% | 0.15% |

Table 8: Subset of Composite Bias

Table 8 shows the rates of *in*consistency between just the first three of those metrics in the sample. With a much bigger sample, these measures do <u>not</u> always signal the direction of bias consistently. The Florida congressional plan above in §6.3 is an example of this.

## Conclusions

The term "partisan bias" means different things in different contexts. Some measures of partisan bias gauge partisan advantage, some packing & cracking, and others partisan symmetry. Different metrics sometimes contradict each other about the direction of bias, sometimes even within the same category of metrics. Both the balance between typical party vote shares and state political geography influence the rate of conflicting signals of bias from different metrics. No subset of the metrics is always internally consistent in its signal of bias. Hence, when assessing the partisan bias of redistricting plans, it is important to be specific about what kind of partisan bias one is interested in and what specific measure or measures one is using to estimate it and why.

## References

[BE24]     Jeffrey T. Barton and Jon X. Eguia. A decomposition of partisan advantage in electoral district maps. Electoral Studies, 92:102871, 2024.

[CDD24]    Sarah Cannon, Daryl DeFord, and Moon Duchin. Repetition effects in a Sequential Monte Carlo sampler. arXiv preprint arXiv:2409.19017, 2024.

[DRA20]    Dave's Redistricting (DRA). https://davesredistricting.org.

[Egu22]    Jon X. Eguia. A measure of partisan advantage in redistricting. Election Law Journal: Rules, Politics, and Policy, 21(1):84–103, 2022.

[KKR20]    Jonathan N. Katz, Gary King, and Elizabeth Rosenblatt. Theoretical foundations and empirical evaluations of partisan fairness in district-based democracies. American Political Science Review, 114(1):164–178, 2020.

[MB15]     Michael McDonald and Robin Best. Unfair partisan gerrymanders in politics and law: A diagnostic applied to six cases. Election Law Journal: Rules, Politics, and Policy, 14(3), 2015.

[MKS+22]   Cory McCartan, Christopher Kenny, Tyler Simko, Shiro Kuriwaki, George Garcia III, Kevin Wang, Melissa Wu, and Kosuke Imai. 50-State redistricting simulations. February 8, 2022. https://doi.org/10.7910/DVN/SLCD3E.

[NR21]     John F. Nagle and Alec Ramsay. On measuring two-party partisan bias in unbalanced states. Election Law Journal: Rules, Politics, and Policy, pages 116–138, 2021. https://doi.org/10.1089/elj.2020.0674.

[Pla]        PlanScore. https://planscore.org.

[Ram23]      Alec Ramsay. Estimating seats–votes partisan advantage. Election Law Journal: Rules, Politics, and Policy, 22(1):67–79, 2023. https://doi.org/10.1089/elj.2022.0031.

[SM15]       Nicholas Stephanopoulos and Eric McGhee. Partisan gerrymandering and the efficiency gap. University of Chicago Law Review, 82:831, 2015. U of Chicago, Public Law Working Paper No. 493. Available at SSRN: https://ssrn.com/abstract=2457468.

[SM18]       Nicholas Stephanopoulos and Eric McGhee. The measure of a metric: The debate over quantifying partisan gerrymandering. Stanford Law Review, 70:1503, May 2018. U of Chicago, Public Law Working Paper No. 672. Available at SSRN: https://ssrn.com/abstract=3077766.

[TPR25]      Kristopher Tapp, Todd Proebsting, and Alec Ramsay. Parameter effects in ReCom ensembles. Preprint, May 2025. Available at arXiv: https://doi.org/10.48550/arXiv.2505.21326.

[Wan16]      Samuel S.-H. Wang. Three tests for practical evaluation of partisan gerrymandering. Stanford Law Review, 68(6):1263–1321, 2016.

[War17]      Gregory S. Warrington. Quantifying gerrymandering using the vote distribution. arXiv preprint arXiv:1705.09393, May 2017.

[War19]      Gregory S. Warrington. A comparison of partisan-gerrymandering measures. arXiv preprint arXiv:1805.12572, April 2019.

## Supplementary Materials

The ensembles and scores are available in the supplemental materials for [TPR25]. The code I used to analyze it for this paper is available in the `partisan_diffs.ipynb` notebook of the `https://github.com/rdatools/rdametrics` GitHub repository.

[end]